\documentclass[prx,showpacs,twocolumn,amsmath,amssymb,floatfix]{revtex4-1}

\usepackage{hyperref}
\usepackage{graphicx}
\usepackage{bm}
\usepackage{color}
\usepackage[applemac]{inputenc}
\usepackage{amssymb} 
 
\newcommand{\sect}[1]{\emph{#1.---}\ignorespaces}     
\newcommand{\edit}[1] {\textcolor{black}{#1}}

\begin{document}  
   
\title{Even-odd effect and Majorana states in full-shell nanowires}

\author{Fernando Pe\~naranda$^{1,2}$, Ram\'on Aguado$^1$, Pablo San-Jose$^1$, Elsa Prada$^2$}
\affiliation{\\
$^1$Instituto de Ciencia de Materiales de Madrid (ICMM), Consejo Superior de Investigaciones Cient\'{i}ficas (CSIC), Sor Juana In\'{e}s de la Cruz 3, 28049 Madrid, Spain. Research Platform on Quantum Technologies (CSIC).\\$^2$Departamento de F\'isica de la Materia Condensada, Condensed Matter Physics Center (IFIMAC) and Instituto Nicol\'as Cabrera, Universidad Aut\'onoma de Madrid, E-28049 Madrid, Spain}

\date{\today} 

\begin{abstract}
Full-shell nanowires (semiconducting nanowires fully coated with a superconducting shell) have been recently presented as a novel means to create Majorana zero modes. In contrast to partially coated nanowires, it has been argued that full-shell nanowires do not require high magnetic fields and low densities to reach a putative topological regime. Here we present a theoretical study of  these devices taking into account all the basic ingredients, including a charge distribution spread across the section of the nanowire, required to qualitatively explain the first experimental results (Vaitiek\.{e}nas {\it et al.}, arXiv:1809.05513). We derive a criterion, dependent on the even-odd occupation of the radial subbands with zero angular momentum, for the appearance of Majorana zero modes. In the absence of angular subband mixing, these give rise to strong zero-bias anomalies in tunneling transport in roughly half of the system's parameter space under an odd number of flux quanta. Due to their coexistence with gapless subbands, the zero modes do not enjoy generic topological protection. Depending on the details of subband mixing in realistic devices, they can develop a topological minigap, acquire a finite lifetime or even be destroyed.
\end{abstract}

\maketitle

Majorana quasiparticles are localized zero-energy excitations, usually arising due to the nontrivial topology of a superconducting bulk~\cite{Read:PRB00,Kitaev:PU01,Hasan:RMP10,Qi:RMP11,Alicea:RPP12,Leijnse:SSAT12,Elliott:RMP15,Aguado:RNC17,Sato:ROPIP17}. Topological protection, together with the non-Abelian braiding statistics of Majoranas, forms the basis of topologically protected quantum computing~\cite{Kitaev:PU01,Nayak:RMP08,Beenakker:ARCMP13}. This prospect has spurred a great deal of efforts in recent years towards their creation and manipulation in various solid-state platforms ~\cite{Mourik:S12,Nadj-Perge:S14,Deng:S16,Zhang:N18,Jeon:S17}. Amongst the most developed is the so-called Majorana nanowire~\cite{Stanescu:JPCM13,Lutchyn:NRM18,Prada:A19}, a proximitized semiconducting nanowire partially coated with a superconductor along its length. The device was designed to realize the Oreg-Lutchyn model~\cite{Oreg:PRL10,Lutchyn:PRL10}, which predicts the emergence of one-dimensional topological superconductivity~\cite{Kitaev:PU01} and protected Majorana zero modes under a strong Zeeman field at low carrier densities. The superconducting coating of the device is limited to some facets of the nanowire to allow depleting the nanowire carrier density with a gate~\cite{Krogstrup:NM15,Sestoft:PRM18,Gill:NL18}, while still preserving a good proximity effect from the superconductor~\cite{Chang:NN15,Gul:NL17}. Majorana-like signatures, e.g., zero-bias anomalies (ZBAs) in transport spectroscopy, have been repeatedly reported in these systems~\cite{Prada:A19}. Despite such promising results, the search continues for alternative platforms or detection schemes~\cite{Shabani:PRB16,Suominen:PRL17,San-Jose:PRX15,San-Jose:PRL14,Bocquillon:NN16,Tiira:NC17,Laroche:NC19,Hell:PRL17,Pientka:PRX17,Finocchiaro:PRL18,Juan:SP19,Fornieri:N19,Ren:N19,Zhang:NC19,Menard:NC19,Heinrich:PSS18,Choi:RMP19,Murani:PRL19,Jack:S19} where Majoranas could also be engineered and manipulated.

\begin{figure}
   \centering
   \includegraphics[width=\columnwidth]{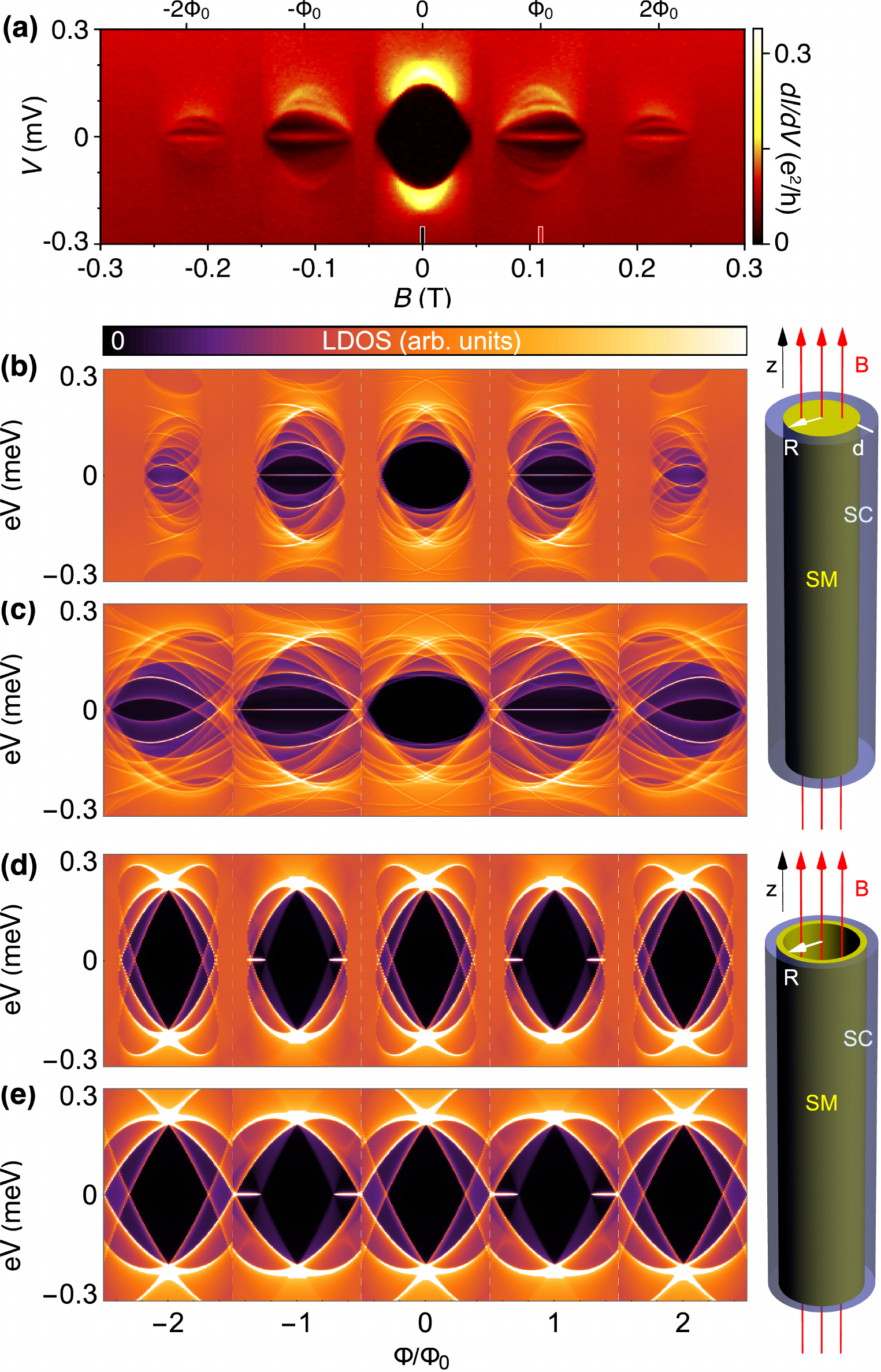}
   \caption{(Color online) (a) Experimental results for differential \edit{tunneling} conductance $dI/dV$ vs bias voltage $V$ and magnetic flux $\Phi$ (or equivalently magnetic field $B$) into a solid-core full-shell superconductor-semiconductor nanowire in the destructive Little-Parks (LP) regime, taken from Ref. \onlinecite{Vaitiekenas:A18}. Zero-bias anomalies, \edit{revealing the presence of zero modes in the LDOS}, are observed in the first LP lobes with $n=\pm 1$ fluxoid around the shell. (b,c) Numerical simulation of the \edit{local density of states (LDOS)} {(in arbitrary units)} in semi-infinite solid-core full-shell nanowires, both in the destructive (b) and weak/moderate (c) LP regime, showing similar phenomenology. Zero modes are absent around integer flux  in simpler hollow-core full-shell nanowires, panels (d,e). Simulation parameters: $\Delta=0.2$ meV, $\alpha =20$ meV nm; (b,c) $\lambda_N=38$ nm, $\lambda_S=35$ nm, $R=80$ nm, and $R^2/(d\xi)=1.72$ (b) and $4.35$ (c); (d,e) $\lambda_N=\lambda_S=61$ nm, $R=43$, $d\approx 0$, and $R/\xi=0.47$ (d) and $0.67$ (e).}
   \label{fig:1}
\end{figure}

A recent experiment~\cite{Vaitiekenas:A18} reported on an innovative type of device, known as a full-shell Majorana nanowire, that appears at first sight to be a minor variation of the Majorana nanowire. The term full-shell refers to the superconducting coating, that is applied on all facets of the nanowire instead of just a few. A full coating prevents external gating of the device as external electric fields can be expected to be totally screened in the nanowire bulk. 
The full-shell geometry, however, also enables new possibilities, particularly the creation of superconducting vortices around the nanowire axis. Under a longitudinal magnetic flux $\Phi$, the order parameter develops an `$n$-fluxoid', $\Delta=|\Delta|e^{in\phi}$, i.e., a winding $n$ of its phase with angle $\phi$ around the nanowire axis, where $n = \lfloor \Phi/\Phi_0\rceil$ is the closest integer to $\Phi$ normalized to the flux quantum $\Phi_0$~\cite{London:50,Brenig:PRL61,Byers:PRL61,Tinkham:04}. Furthermore, the Little--Parks (LP) effect \cite{Little:PRL62,Tinkham:04} arises, whereby the superconducting gap $|\Delta|$ becomes suppressed (even completely in the `destructive' regime \cite{De-Gennes:CRAS81,Liu:S01}) around half-integer flux.  It was found experimentally \cite{Vaitiekenas:A18} that, in the presence of an odd-$n$ fluxoid, a Majorana-like ZBA arises in the nanowire at magnetic fields much smaller than in partial-shell devices. It was furthermore found to remain robust for any magnetic flux throughout the `first lobe' centered around $\Phi/\Phi_0\approx 1$, see Fig. \ref{fig:1} (a).

As shown in Ref.~\cite{Lutchyn:A18}, a hollow version of the nanowire can be mapped analytically to the Oreg-Luthyn model, which could then sustain topologically protected states without the need of a Zeeman field, an essential ingredient in the original Oreg-Lutchyn proposal~\cite{Oreg:PRL10,Lutchyn:PRL10}. In such case, however, the corresponding Majorana zero modes only survive near the edge of the odd lobes, see Figs. \ref{fig:1} (d,e), but not near the center of the lobe, unlike in the experiment. The theory analysis also showed that the more realistic case of a solid full-shell nanowire can also exhibit a topological phase. Its parameter window, however, was found to be very small and restricted to low densities, at least in the case of a pristine nanowire with perfect circular symmetry, and would require fine control of its density to be realized. Away from this small window, it was shown that the system is gapless, due to the presence of ungapped subbands with higher angular momentum components. An open question thus remains as to the nature of the experimental ZBAs, that, surprisingly, required no fine tuning of gates or field.

In this work we address this question by studying the spectral properties of more general full-shell nanowires {with a solid core, generalizing previous results to the realistic case in which charge density is spread across nanowire section. We find that unprotected but strong Majorana-like ZBAs arise from the sector with lowest angular momentum $m_j=0$, embedded in a gapless $m_j\neq 0$ background. Their emergence results from a nontrivial topology of the $m_j=0$ sector when the occupation of the corresponding normal-state radial subbands is odd. We compute the system's phase diagram, which clearly reveals this even-odd effect, with ZBAs present throughout a substantial fraction of parameter space. We further demonstrate that ZBAs persist across odd lobes. Our spectroscopy simulations shows a marked similarity to the experimental observations without the need of fine tuning.}
The resulting Majorana states are however unprotected against general subband-mixing perturbations (from, e.g., interface disorder or a noncircular nanowire section or shell), since they coexist with a gapless background, {as also noted in Ref.} \onlinecite{Lutchyn:A18}. We explore {here} their fate in the presence of angular mode mixing. Depending on the mixing details, we find a variety of possible behaviors, including the development of a trivial or a nontrivial gap, a splitting or a broadening of the zero mode into a delocalized quasibound Majorana state. We conclude by commenting on possible alternative scenarios for the observations.

\sect{Model}
We first develop the simplest description of a solid semiconducting nanowire of radius $R$, oriented along the $z$ direction, and fully coated with a conventional superconductor of thickness $d$. The Fermi energies of the two materials are denoted by $\mu_N$ and $\mu_S$ respectively, with $\mu_S\gg\mu_N$. The associated Fermi wavelengths are denoted by $\lambda_{N,S} = \hbar/\sqrt{2m^*\mu_{N,S}}$, with $m^*$ the effective mass (assumed uniform for simplicity). When the nanowire core is contacted to the superconducting shell, $\mu(r)$ {will in general acquire self-consistent screening corrections. We assume instead} the simple approximation $\mu(r<R)=\mu_N,\, \mu(r>R)=\mu_S$. {While the chemical potential is piecewise contact, the resulting charge density is not, acquiring a nontrivial radial profile that affects the \edit{local density of states (LDOS) measured} by a tunnel probe}. Similarly, {we assume} $|\Delta(|r|<R)|=0,\, |\Delta(r>R)| = |\Delta|$. The dependence of $|\Delta|$ with flux $\Phi$ is incorporated from the LP Ginzburg-Landau theory results, see Appendix \ref{appendixA}, whose high accuracy has been recently established~\cite{Vaitiekenas:PRB20}. The relevant spin-orbit Rashba coupling inside the nanowire is radial, $\bm{\alpha}(r)\parallel\hat{\bm{r}}$, and is much smaller in the superconductor than in the semiconductor. We approximate $\bm{\alpha}(r<R) = \alpha \frac{r}{R}\,\hat{\bm{r}},\,\bm{\alpha}(r>R) =  0$~\cite{Antipov:PRX18,Mikkelsen:PRX18,Escribano:PRB19}.
The section of the nanowire is assumed circular for the moment, so that subbands have a well defined total angular momentum $m_j$. The three-dimensional Nambu Hamiltonian for this model can be written in cylindrical coordinates as
\begin{eqnarray}
\label{Hamiltonian}
H &=& \left[\frac{(\bm{p}+e\bm{A})^2}{2m^*} -\mu(r) + \bm{\alpha}(r)\cdot\bm{\sigma}\times\left(\bm{p} + e\bm{A}(r)\right)\right]\tau_z\nonumber \\
&& + \sigma_y\tau_y|\Delta(r)| e^{in\phi}, \label{H}
\end{eqnarray}
where $\sigma_i$ are Pauli matrices for spin, and $\tau_i$ for the electron-hole sectors. The magnetic flux is incorporated through the $n$-fluxoid in the pairing term and through the axial gauge field $\bm{A}(\bm{r}) \approx \frac{r\Phi}{2\pi R^2}\hat{\bm{\phi}}$, where $\hat{\bm{\phi}}$ is the axial unit vector in cylindrical coordinates, $\Phi=\pi B R^2$ is the flux and $B$ is the magnetic field along the $z$ direction.

Due to the axial symmetry of the model, the above $H$ can be decomposed into decoupled sectors with different total angular momentum $m_j$~\cite{Lutchyn:A18}. By discretizing the resulting $H_{m_j}$ into a one-dimensional semi-infinite tight-binding Hamiltonian along the $z$ direction, we can compute the total \edit{LDOS at the end of the nanowire} as a sum of different $m_j$ contributions. \edit{Experimentally, the LDOS is measured with the $dI/dV$ conductance through a tunnel probe coupled to the end of the wire} \footnote{{This assumes a model for the tunnel contact that couples the reservoir to all $m_j$ modes equally, as would correspond to a dirty contact limit.}}. At small bias voltage $V$ and temperature $T$, the tunneling $dI/dV$ is an approximate measure of the nanowire \edit{LDOS at} energy $eV$. \edit{An LDOS ZBA in this context thus refers to the existence of a zero-energy mode} \footnote{\edit{In the presence of a zero-energy Majorana bound state, the only difference between the LDOS ZBA and the $dI/dV$ ZBA is that while the former is technically a delta function, the latter is a peak with a universal zero-temperature height of $2e^2/h$ and a width given by the tunnel coupling. Finite temperature $T$ broadens and lowers the peak, but it remains visible as long as $k_BT<|\Delta|$).}}. We compute the \edit{LDOS} using the Green's function formalism for one-dimensional, semi-infinite conductors \cite{Economou:83,Datta:97,MathQ}.

\sect{Results and discussion} In Fig. \ref{fig:1} we show a comparison between the experimental $dI/dV$ as a function of bias $V$ and flux $\Phi$ (destructive LP regime) and our \edit{LDOS} simulations (both in the weak and destructive LP regimes). The LP regime is mainly controlled by the ratio $R^2/(d\xi)$ between nanowire radius $R$ and superconductor coherence length $\xi$ and thickness $d$; a large ratio giving weak LP (see Appendix \ref{appendixA}). We also present the corresponding simulation for a hollow nanowire, wherein the semiconductor is confined to a very thin shell $R-\delta<r<R$ with $\delta\ll R$ (see sketch). The experiment shows a strong ZBA clearly visible throughout the odd $n=\pm 1$ LP lobes, including $\Phi/\Phi_0= 1$ and its vicinity. This feature is reproduced by LDOS simulations in the solid nanowire model. A ZBA across the lobe like in the experiment, however, is not reproduced by the hollow nanowire model, as already demonstrated in Ref. \onlinecite{Lutchyn:A18}.

\begin{figure}
   \centering
   \includegraphics[width=\columnwidth]{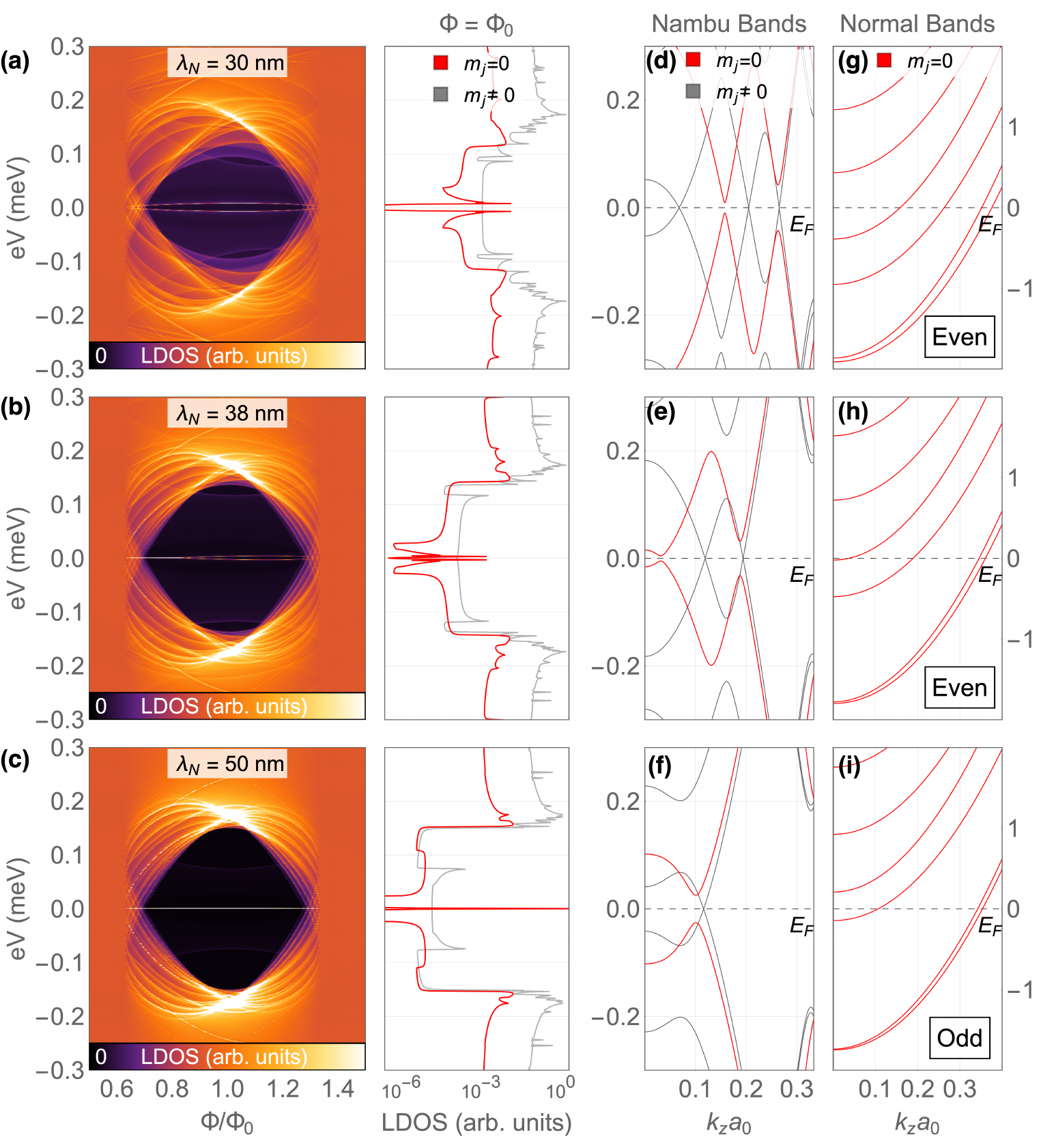}
   \caption{(Color online) (a-c) \edit{LDOS at the end of} a semi-infinite solid-core, full-shell nanowire as a function of \edit{energy $eV$} and flux $\Phi$ in the first LP lobe (destructive regime). {The Fermi wavelength of the superconducting shell $\lambda_S$ is fixed and the semiconducting core's $\lambda_N$ increases from (a) to (c).} Log-scale line cuts {of the \edit{LDOS}} at $\Phi=\Phi_0$ {are shown} on the right, resolved by sectors of different angular momentum $m_j$. {The line cuts show that black regions in the density plots have a small nonzero \edit{LDOS} background coming from gapless $m_j\neq 0$ subbands, not visible with the chosen color scaling (as also happens in the experiment of Ref. \onlinecite{Vaitiekenas:A18}).} Note the appearance of a quasibound zero mode throughout the lobe in (b,c) that was split in (a). The contributions to the \edit{LDOS} from sectors with different $m_j$ show that the zero mode corresponds to the $m_j=0$ sector {(in red)}. It arises as one $m_j=0$ subband undergoes an inversion at $k_z=0$, as shown in the Nambu bandstructure in panels (d-f), thus becoming topologically nontrivial when considered on its own. The emergence of the $m_j=0$ Majorana zero mode correlates with an odd occupancy of the $m_j=0$ radial subbands in the normal phase, panels (g-i). Parameters are as in Fig. \ref{fig:1}(b) except for $R = 100$ nm and $\lambda_S=24$ nm.}
   \label{fig:2}
\end{figure}

\begin{figure}
   \centering
   \includegraphics[width=.9\columnwidth]{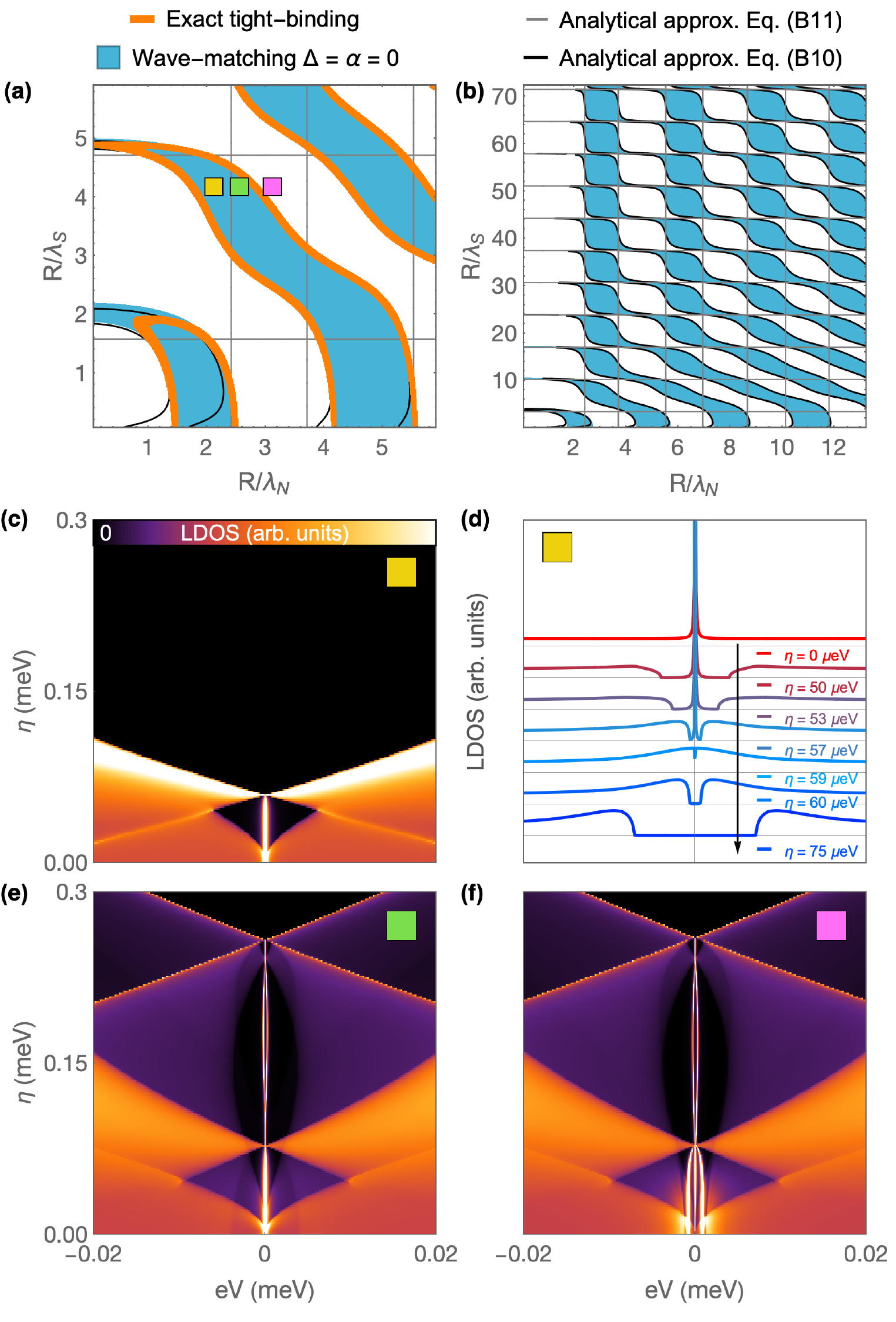}
   \caption{(Color online) (a,b) Phase diagram of a $\Phi=\Phi_0$ solid-core, full-shell nanowire of radius $R$, vs $R/\lambda_{N,S}$, where $\lambda_{N,S}$ are Fermi wavelengths in the semiconductor core and superconductor shell, respectively. Panel (a) focuses on low densities while (b) shows a wider range. The thick orange lines in (a) mark the boundaries of regions with a topologically nontrivial $m_j=0$ subband with Majoranas. These are computed using exact tight-binding simulations with finite $\Delta$. Blue regions in (a,b) correspond to an odd occupancy of the $m_j=0$ normal-phase radial subbands, computed using wavematching at the core-shell boundary for $\Delta=\alpha=0$. Black and gray lines correspond to two analytical approximations for the even-odd boundaries; see Eqs. \eqref{secondorder} and \eqref{zeroorder}. (c-f) \edit{LDOS} as a function of \edit{energy $eV$} and subband-mixing strength $\eta$, starting at different points in the phase diagram, colored squares in (a). {Note that the color scale is zoomed in around zero conductance with respect to Figs. \ref{fig:1} and \ref{fig:2} to resolve the small background \edit{LDOS}} (d) line cuts of (c) that emphasize the opening and subsequent band inversion of a minigap induced by mixing, which in our model eventually results in the destruction of the Majorana state. Parameters: (a) $R=100$ nm, $d=100$ nm and $\Delta =0.2$ meV; $\alpha = 10$ meV nm for orange curve; (b) $R=65$ nm, $d=28$ nm.}
    \label{fig:3}
\end{figure}

Due to the nongateable nature of the full-shell devices, $\mu_N$ and to some extent also $\mu_S$ are unknown. It is thus important to establish when ZBAs arise as a function of these two quantities.  To this end we first analyze the solid nanowire \edit{LDOS} for fixed $\mu_S$ and decreasing $\mu_N$. This corresponds to fixing $\lambda_S$ and increasing $\lambda_N$. Simultaneously we compute the $m_j$-resolved bandstructure at $\Phi/\Phi_0=1$, both in the superconducting and the normal phase. The combined results are shown in Fig. \ref{fig:2}. We find a trivial phase with split ZBAs [panel (a)] that transitions to a nontrivial phase with an unsplit ZBA [panel (c)] corresponding to a Majorana bound state localized at the tunneling contact. This happens whenever an $m_j=0$ Nambu subband, in red in panels (d-f), becomes inverted. 

The topological phases accurately correlate, in the limit of $\Delta\ll\mu_S,\mu_N$, with an odd occupation of the normal-state $m_j=0$ radial subbands, panels (g-i). These normal subbands are spread throughout the inner core and the outer shell of the nanowire, so that the precise transition point depends on both $\lambda_N$ and $\lambda_S$. It also depends weakly on $\Phi$. This is demonstrated in Fig. \ref{fig:2} (b), where $\lambda_N$ and $\lambda_S$ are tuned to the vicinity of an even-odd transition. There, $\Phi/\Phi_0\lesssim 1$ has odd occupancy and a ZBA exactly at zero, while for $\Phi/\Phi_0\gtrsim 1$ the occupancy is still even, and the ZBA exhibits a weak splitting. Such $\Phi$-dependence within odd lobes is however quite weak in practice.

{The phase diagram of the model at fixed $\Phi=\Phi_0$ is shown in Fig. \ref{fig:3} (a), where we compare the normal-phase odd-occupancy criterion (blue regions), the emergence boundary of Majorana zero modes (orange lines), and two analytical approximations for the latter, Eqs. \eqref{secondorder} and \eqref{zeroorder} of Appendix \ref{appendixB} (black and gray lines). We find that the unsplit $m_j=0$ Majoranas are a common occurrence, occupying essentially half of the phase diagram. No fine tuning is thus necessary to achieve such a phase, which could explain why the ungateable experimental nanowires are likely to show this phenomenology.}


The background of gapless $m_j\neq 0$ modes, represented in gray in the second column of Fig. \ref{fig:2}, provides a continuum of excitations for the $m_j=0$ Majorana to couple to or decay into \footnote{{Note that the presence of this small background conductance around the zero mode is not evident in the $dI/dV$ density plots of Fig. \ref{fig:1} due to the scale of the color bar, which in our simulations has been chosen to match the experimental results of Fig. \ref{fig:1}(a). However, both in theory as well as in the experiment, the conductance around zero-bias is not zero inside the first lobe, despite what the almost black color may suggest.}}. As a result, the Majorana zero mode does not enjoy generic topological protection in full-shell nanowires, as mode mixing can potentially destroy it. To understand how, we have performed simulations using a minimal model for angular mode mixing, in line with our nanowire model; see Appendix \ref{appendixC} for implementation details. A single parameter $\eta$ controls the strength of $m_j$ mixing, with all preceding results corresponding to $\eta=0$. In Figs. \ref{fig:3} (c-e) we show the evolution of the \edit{LDOS} in two topological points of the phase diagram [yellow and green in (a)] as a uniform $\eta$ is increased throughout the semi-infinite nanowire. The \edit{LDOS} first develops a small topological minigap (black background around the ZBA), which then closes and reopens at a critical value of $\eta$, destroying the ZBA (c). The corresponding behavior starting within a trivial phase [pink in (a)] is shown in (f). We see that considerably complex evolutions with $\eta$ may arise, including intermediate phases with additional pairs of zero modes. With a spatially non-uniform $\eta$ we even see mode broadening into a Majorana quasibound state; see Appendix \ref{appendixC}. We find that within our model mode mixing invariably ends up by destroying all zero modes at large $\eta$, which suggests that strongly asymmetric or disordered nanowires will be poor candidates to exhibit this type of zero modes.

\sect{Conclusion} We have established the minimal ingredients necessary to model and explain the subgap tunneling $dI/dV$ phenomenology of full-shell superconductor-semiconductor nanowires of recent experiments~\cite{Vaitiekenas:A18}. The hollow-core version never shows ZBAs throughout a full LP lobe. It is necessary to consider solid-core nanowires with a nonzero charge density throughout the full nanowire section to obtain ZBAs similar to the experiment. We showed that these emerge for odd normal-state occupation of the radial $m_j=0$ subbands. We have mapped analytically and numerically this even-odd effect in the emergence of ZBAs at odd LP lobes throughout the full phase diagram of the system's model, and established the connection between the ZBAs to topologically unprotected $m_j=0$ Majorana zero modes. We have found that, while they are not a signature of robust topologically protected zero modes, unsplit ZBAs should be a common occurrence in these devices, occupying roughly half of their microscopic parameter space at zero temperature. We also found that the effect of angular subband mixing on the Majoranas is quite complex, ranging from topological minigap opening to mode splitting or broadening, but always ends up by destroying the Majorana states at sufficiently strong mixing.

While the physical picture presented here is qualitatively consistent with the first batch of experimental results, it also implies that the emergence of Majoranas in these devices is hard to predict and control, owing to its dependence on the even-odd occupation of $m_j=0$ radial subbands, and their mixing with other $m_j\neq 0$. Another important prediction is the lack of a ZBA within even-$n$ lobes. Given the current resolution, it is not clear from the available data in \cite{Vaitiekenas:A18} whether the $n=2$ hosts a ZBA or a low-lying split resonance. Thus, other common ZBA-generating mechanisms, such as smooth confinement at the tunnel contact and unintentional quantum dot formation, should not be discarded as alternative scenarios in future studies.

\sect{Note added}
While this work was being reviewed for publication, Refs. \onlinecite{Vaitiekenas:A18} and \onlinecite{Lutchyn:A18} were merged and published as a single paper \cite{Vaitiekenas:S20}.

\acknowledgements

Research supported by the Spanish Ministry of Science, Innovation and Universities through Grants No. FIS2015-65706-P, No. FIS2015-64654-P, No. FIS2016-80434-P, No. PCI2018-093026, and No. PGC2018-097018-B-I00 (AEI/FEDER, EU), the Ram\'on y Cajal program Grant No. RYC-2011-09345, the Mar\'ia de Maeztu Program for Units of Excellence in R\&D (Grant No. MDM-2014-0377), and the European Union's Horizon 2020 research and innovation program under FETOPEN Grant Agreement No. 828948. We also acknowledge support from CSIC Research Platform on Quantum Technologies PTI-001.

\appendix

\section{Destructive Little-Parks oscillations}
\label{appendixA}

{A hollow superconducting cylinder has a single-valued complex order parameter $\Delta(\phi)$ around its axis, where $\phi$ is the azimuthal angle. Assuming its modulus is never zero, the phase of $\Delta$ can have only an integer winding, i.e., $\Delta(\phi)=|\Delta|e^{in\phi}$ where $n$ is an integer and $\phi\in[0,2\pi)$. This is known as fluxoid quantization. In a hollow superconducting cylinder longitudinally threaded by a magnetic field, this condition leads to a periodic modulation of the self-consistent $|\Delta|$ (and hence of the superconducting critical temperature $T_C$) as a function of the applied flux ($\Phi$). The maximal $T_C$ reduction is located at half-integer normalized flux ($\bar{\Phi} = \Phi/\Phi_0 = m + 1/2$ for integer $m$). This is known as the Little-Parks (LP) effect predicted in Refs. \onlinecite{London:50,Brenig:PRL61,Byers:PRL61} and first observed in Ref. \onlinecite{Little:PRL62}.}

{An additional pair breaking mechanism must be taken into account in superconducting cylinder shells with small radius $R$ and thickness $d$ as compared to the zero-temperature coherence length $\xi$. Such pair breaking stems from the flux-induced supercurrents circulating the shell, which may lead to drastic deviations from the original LP prediction for large $R$, $d$. In the small $R/\xi\ll 1$ regime (which is the relevant one in the experiments of Refs. \onlinecite{Vaitiekenas:A18,Vaitiekenas:PRB20}) the modulation of $T_C$ results in a full suppression of superconductivity in finite flux regions and, therefore, metallic and superconducting phases alternate with the applied flux at $T=0$. This is the so-called strong (or destructive) Little-Parks regime (SLP) \cite{De-Gennes:CRAS81, Liu:S01}.}

{The effect of the flux-induced supercurrent as a Cooper pair breaking mechanism is formally equivalent to that caused by paramagnetic impurities \cite{Gorkov:JETP}, as shown in Refs. \onlinecite{Dao:PRB,Schwiete:PRB10}. Thus, in order to find a formula for $T_C(\Phi)$ in the SLP regime, we use a mean-field approach in close analogy to the standard theory of the microscopic time-dependent Ginzburg-Landau (TDGL) description of a disordered superconductor with paramagnetic impurities, as was done in Ref. \onlinecite{Schwiete:PRB10}. This involves starting from the TDGL functional integral in the saddle-point approximation and computing the poles of the two-particle Green's function fluctuation propagator:
$ \mathcal{L}(p,p',q) = \langle \mathcal{T_\tau}(\psi_{p+q,\sigma}\psi_{-p,-\sigma},\psi^\dagger_{p'+q,\sigma'}\psi^\dagger_{-p',-\sigma'})\rangle$ 
up to zeroth order in its Dyson expansion, where $\psi^\dagger_{p,\sigma}$ is a fermionic creation field operator in the Heisenberg representation with momentum $p$ and spin $\sigma$. This results in the following condition (see Ref. \onlinecite{Larkin:Oxford} for a comprehensive derivation) from which $T_C(\bar{\Phi})$ can be readily calculated:
\begin{eqnarray}
0 &=& (\nu \mathcal{L})^{-1}(n,\omega=0) \\ 
&=& \ln \left( \frac{T_C}{T_C^0} \right) + \Psi \left( \frac{1}{2} + \frac{\Lambda_n(\Phi) }{2\pi T_C(\bar{\Phi})} \right) - \Psi \left( \frac{1}{2} \right),\nonumber
\end{eqnarray}
where $\nu$ is density of states at the Fermi level, $T_C^0$ is the critical temperature of the bulk material at zero flux, $\Psi$ is the digamma function, and $\Lambda_n$ is the pair breaking function corresponding to hollow superconducting cylinders. The latter is obtained by solving the GL equations in the presence of impurities \cite{Groff:PR}. As a function of the fluxoid winding number $n$, it reads
\[
\Lambda_n(\bar{\Phi}) = \frac{T_C^0}{\pi} \frac{\xi^2}{R^2}\Big[ 4(n-\bar{\Phi})^2 +\frac{d^2}{R^2} \left( \bar{\Phi}^2 + \left( \frac{1}{3} + \frac{d^2}{20R^2}\right) \right)n^2  \Big].
\]
}


\section{Derivation of the normal-phase odd-occupancy criterion in odd-flux lobes}
\label{appendixB}

In this Appendix we tackle the problem of computing the occupation of the $m_j=0$ subbands in the normal state, whose parity is an approximate criterion for the existence of Majorana zero modes. This approximation is shown to be valid in the main text throughout odd-flux LP lobes, in the Andreev limit ($\Delta\ll\mu_S, \mu_N$) and in the absence of subband mixing. We perform the calculation using wavematching techniques at the semiconductor-superconductor boundary, and derive analytical approximations for the odd-parity regions in parameter space.

At zero temperature, the occupation of the $m_j=0$, normal-state, radial subbands changes when one of them crosses the Fermi level. This always happens at $k_z=0$ in odd-flux lobes. Therefore, we look for the solutions of
\begin{equation}
H_{m_j=0} (r) \Psi (r) = 0,
\end{equation}
where $H_{m_j=0}$ is the $m_j=0$ projection of the normal-state (i.e., $\Delta = 0$) Hamiltonian of Eq. \eqref{Hamiltonian}. We write it as a piecewise-constant combination of nanowire $H_{N}(r)$ and shell $H_{S}(r)$ Hamiltonians as~\cite{Lutchyn:A18}
\begin{eqnarray}
H_{m_j=0}(r) = H_{N}(r) \theta(R-r) + H_S (r) \theta (r-R),
\end{eqnarray}
with
\begin{eqnarray}
H_N(r) &=& \left(\frac{-1}{2  m^* r}  \partial_r r \partial_r - \mu_N \right) \sigma_0 \nonumber \\
&& -\frac{1}{8  m^* r^2} \left( \sigma_z  + \left( 1 - \bar{\Phi} \frac{r^2}{R^2} \right)^2 \sigma_0  \right)^2 \nonumber  \\
&& + \frac{\alpha}{2 r} \sigma_z \left( \sigma_z  + \left( 1 - \bar{\Phi} \frac{r^2}{R^2} \right) \sigma_0 \right)
\label{s0}
\end{eqnarray}
and
\begin{eqnarray}
H_S(r) &=& \left(  \frac{-1}{2 m^* r} \partial_r r \partial_r - \mu_S \right) \sigma_0 \nonumber \\ 
&& - \frac{1}{8 m^* r^2} \left( \sigma_z + ( 1 - \bar{\Phi}) \sigma_0 \right)^2. 
\end{eqnarray}
$\bar{\Phi}$ is the externally applied magnetic flux normalized by the flux quantum, i.e.,  $\bar{\Phi} \equiv \frac{\Phi}{\Phi_0}$.
For the purpose of finding the occupation boundaries we take the $\alpha = 0$ limit in Eq. (\ref{s0}). This is a good approximation, since occupation boundaries occur by band inversions at $k=0$, where spin-orbit coupling is zero. Our steplike model for the chemical potential in the radial direction reads
\begin{equation}
    \mu(r)=
    \begin{cases}
      \mu_{N}, & \text{if}\ r\leq R \\
      \mu_{S}, & \text{otherwise}.
    \end{cases}
\end{equation}

We proceed by wave matching the zero energy solutions for the core and shell regions at $r=R$. The problem reduces to solving a system of four uncoupled equations corresponding to the semiconducting and superconducting  bi-spinors:  $\phi^N=(\phi^N_{\uparrow},\phi^N_{\downarrow})$ and $\phi_S=(\phi^S_{\uparrow}, \phi^S_{\downarrow})$, respectively. These read
\begin{eqnarray*}
0 &=& \phi^N_{\uparrow}(r)'' + \frac{\phi^N_{\uparrow}(r)'}{r} \nonumber \\
&& +\phi^N_{\uparrow}(r) \Bigl(  -\frac{1}{r^2} +( 2 m^*\mu_N  +\frac{\bar{\Phi}}{R^2})   - \frac{\bar{\Phi}^2 r^2}{4 R^4}\Bigr),\\
0 &=& \phi^N_{\downarrow}(r)''+ \frac{\phi^N_{\downarrow}(r)'}{r}+ \phi^N_{\downarrow}(r)\left( 2 m^* \mu_N - \frac{\bar{\Phi}^2 r^2}{4 R^4} 
 \right),\\
0 &=& \phi^S_{\uparrow}(r)'' + \frac{\phi^S_{\uparrow}(r)'}{r}\nonumber \\
&&+\phi^S_{\uparrow}(r) \left(2 m^* \mu_S- \frac{1}{4 r^2} (4-2\bar{\Phi}-\bar{\Phi}^2) \right),\\
0&=&\phi^S_{\downarrow}(r)''+\frac{\phi^S_{\downarrow}(r)'}{r}+\phi^S_{\downarrow}(r)\left(  2 m^*\mu_S  -\frac{1}{4 r^2}\right).
\end{eqnarray*}
The solutions to the above set of equations which satisfy regularity at the origin are given by 
\begin{eqnarray}
\phi^N_{\uparrow} (r) &=& C_{N \uparrow} e^{-\Phi r^2/(4R^2)} r^{-1} L^{(-1)}_{\nu_1} \left(\frac{\bar{\Phi}}{2} \frac{r^2}{R^2}\right), \\
 \phi^N_{\downarrow}(r) &=&C_{N \downarrow} e^{-\Phi r^2/(4R^2)} L^{(1)}_{\nu_2}\left( \frac{\bar{\Phi}}{2} \frac{r^2}{R^2} \right), \\
 \phi^S_{\uparrow}(r) &=& C_{S \uparrow}^a  J_{\nu_\Phi} \left(r/\lambda_S \right) +  C_{S \uparrow}^b  Y_{\nu_\Phi} \left(r/\lambda_S \right), 
\label{s1}\\
 \phi^S_{\downarrow}(r) &=& C_{S \downarrow}^a J_{1/2} \left(r/\lambda_S \right) +C_{S \downarrow}^b Y_{1/2}\left(r/\lambda_S \right),
\label{s2}
\end{eqnarray}
with
\begin{eqnarray*}
\nu_i &=& \frac{3}{2 } +  \frac{R^2}{2\bar{\Phi} \lambda_N^2} - i, \hspace{1cm}(i = 1, 2)\\
\nu_\Phi &=& \sqrt{4-2\bar{\Phi}+\bar{\Phi}^2}/2.
\end{eqnarray*}
$C_x^y$ are arbitrary constants, $ \lambda_{N/S}^{-1} = \sqrt{ 2 m^* \mu_{N/S} } $  are the Fermi wavelengths in the semiconducting and superconducting regions, respectively, $L_\nu^{(a)}(x)$ are the generalized Laguerre polynomials, and $J_\nu(x)$ and $Y_\nu(x)$ are the Bessel functions of first and second kind, respectively.

By imposing continuity of the solutions and their derivatives at $r=R$, and by requiring $u_S(r)$ to vanish at the outermost radius ($r=R+d$), we obtain a set of two independent transcendental equations ($\pi_1=0$ and $\pi_2=0$) for each spin components that relates parameters $\mu_N,\mu_S$, and $\bar{\Phi}$ at the occupation boundaries.
For the sake of conciseness we omit the explicit and lengthy expressions, which correspond to the boundaries of the blue regions in Figs. 3 (a,b) of the main text.\\

A more compact formula for the {occupation boundaries} can be derived by expanding to second order in $\frac{\lambda_S}{R}$ in the asymptotic series of Eqs. (\ref{s1}-\ref{s2}) (since $\lambda_S\ll R$ is the experimentally relevant situation) and to zero order in the expansion of $\pi_{1,2}$  around $\Phi=\Phi_0$.
Under such approximations, occupation boundaries satisfy
\begin{eqnarray}\label{secondorder}
 \Big(   2 f L^{(-1)}_{\rho_1}(1/2)-\frac{\lambda_S}{R} \big(L^{(0)}_{\rho_2}(1/2)+3 L^{(-1)}_{\rho_1}(1/2)\Big)\Big) &\times&  \nonumber\\
\Big(2 f L^{(0)}_{\rho_2}(1/2)-\frac{\lambda_S}{R} \big(L^{(0)}_{\rho_2}(1/2)+2 L_{\rho_3}^{(1)}(1/2)  \Big) \Big) & = &0,\nonumber\\
\end{eqnarray}
with
\begin{eqnarray*}
f =& -&\cot\left(\frac{d}{\lambda_S}\right)\\
& +& \frac{\lambda_S}{R} \left( 1 - \frac{\lambda_S}{R} s \cot\left(\frac{d}{\lambda_S} \right) \right) \left(m+s \cot\left(\frac{d}{\lambda_S}\right)^2\right),\nonumber\\
m&=&\frac{1}{8} \left( \frac{4R+d}{R+d}\right),\\
s&=&\frac{1}{8} \left( \frac{d}{R+d}\right),\\
\rho_i &=& \frac{3}{2} +  \frac{R^2}{2\lambda_N^2} - i.\hspace{1cm}(i = 1, 2, 3)\\
\end{eqnarray*}
Note that replacing $=0$ by $>0$ in Eq. (\ref{secondorder}) above actually selects the regions with odd normal occupation. Equation (\ref{secondorder}) is plotted in black in Figs. 3(a,b) of the main text.

Alternatively, expanding $\pi_{1,2}$ up to leading order in both $\lambda_S/\lambda_N$ and $\lambda_S/R$, reduces to
\begin{eqnarray}\label{zeroorder}
&& \cos\left(\frac{d}{\lambda_S}\right)L^{(-1)}_{\nu_1}\left(\frac{\bar{\Phi}}{2}\right)L^{(0)}_{\nu_2}\left(\frac{\bar{\Phi}}{2}\right) = 0.
\end{eqnarray}
which corresponds to the square mesh plotted in gray in Figs. 3(a,b). Despite its simplicity, the above equation captures quite well the essence of the even-odd effect in the ZBA of our full-shell nanowire model.

\section{Fate of the $m_j=0$ Majorana under interband mixing}
\label{appendixC}

\begin{figure}
   \centering
   \includegraphics[width=0.7\columnwidth]{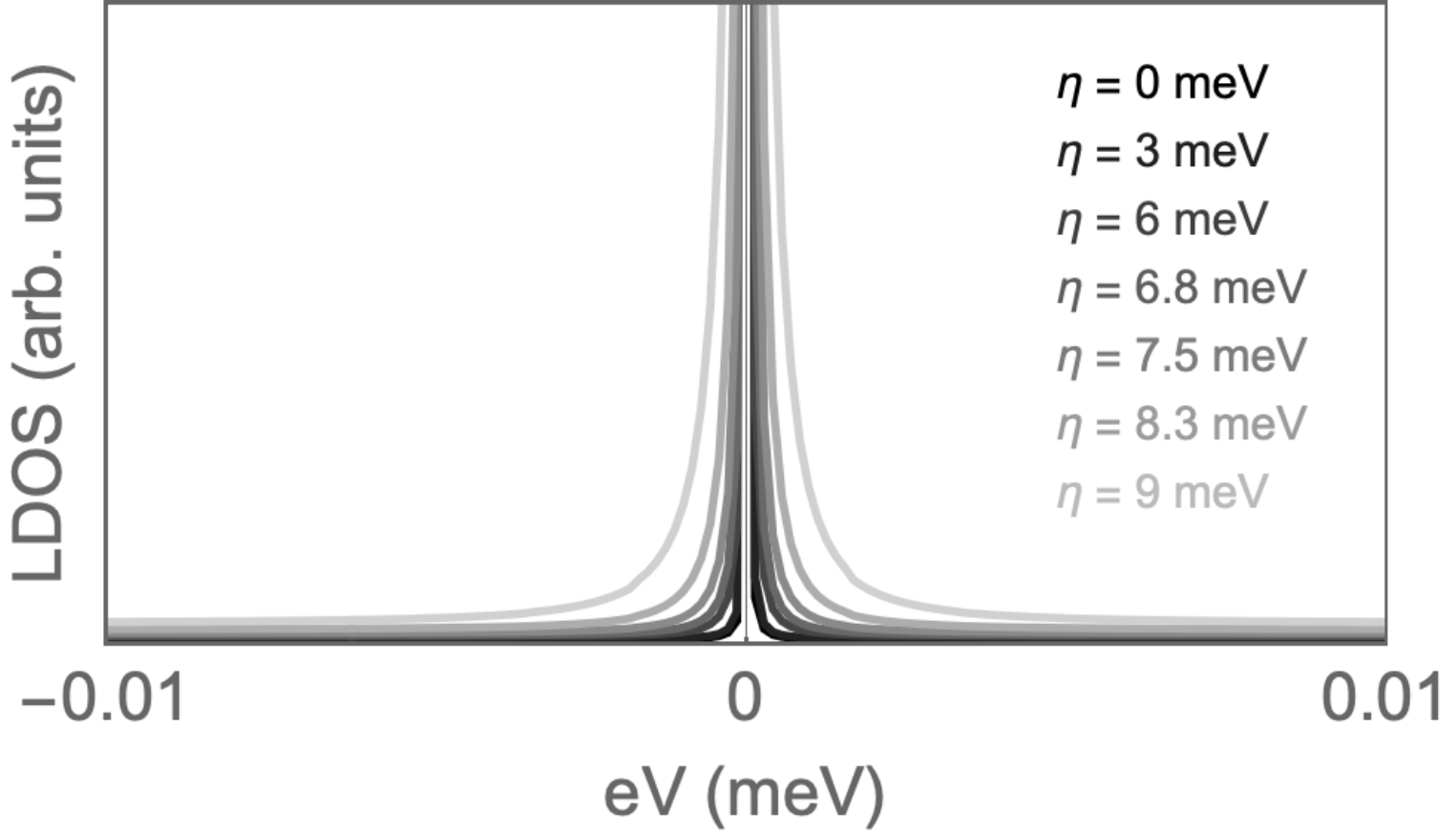}
   \caption{(Color online) The \edit{LDOS at} the end of a semi-infinite full-shell nanowire at $\Phi=\Phi_0$, truncated to the $m_j=0, \pm 1$ subspace, as a function of the coupling $\eta$ between $m_j$ bands. The coupling is restricted to within 10 nm of the end of the nanowire, where the termination of the superconducting shell exposes the semiconductor to external perturbations. The $m_j=0$ Majorana state is coupled by the local $\eta$ to the gapless $m_j\neq 0$ bands in the nanowire bulk, which leads to broadening and decay in the limit of a semi-infinite nanowire.}
   \label{fig:4}
\end{figure}

The analysis of the full-shell nanowire based on decoupled $m_j$'s is valid in the idealized limit of nanowires with perfect cylindrical symmetry. Any {perturbation $\hat V_\eta$}, such as a non-circular section, disorder in the semiconductor, in the superconductor shell or contact, or produced by the presence of a substrate, should be expected to break the assumption of decoupled $m_j$'s to some degree, {as was noted in Ref. \onlinecite{Lutchyn:A18}}. To assess the likelihood of observing the $m_j=0$ ZBA phenomenology connected to Majorana states, we compute and analyze the local density of states (LDOS) under an increasing coupling {$\eta = \langle \phi_{\pm 1}|\hat V_\eta|\phi_0\rangle$} between a small set of angular momenta $m_j=0,\pm 1$ at $\Phi=\Phi_0$ (adding higher bands does not change the qualitative results). As we discussed in the main text, this simplified model is enough to produce a very rich set of possible evolutions of the $m_j=0$ Majoranas, eventually leading to its destruction at strong enough mixing.

The interband mixing is introduced as a uniform coupling $\eta$ between $m_j=0$ and $m_j=\pm 1$. We first assume {$\hat V_\eta$ (and hence also $\eta$)} is independent of position. With a finite $\eta$, the LDOS is no longer decomposable into different $m_j$ contributions. In Figs. 3 (c-f) of the main text we present the total LDOS at $\Phi=\Phi_0$ for increasing $\eta$, starting from different points in the phase diagram of Fig. 3 (a). In (c) we see the simplest possibility. Starting in a nontrivial configuration with one zero mode, a small $\eta$ creates a minigap in the $m_j\neq 0$ subbands by making these modes susceptible to superconducting pairing at zero energy, which otherwise only affects the $m_j=0$ sector. The minigap acts as a proper topological gap, and protects the Majorana much as in conventional Oreg-Lutchyn nanowires. As $\eta$ is increased further, however, the minigap eventually closes and reopens as a trivial gap, destroying the Majorana. 

Starting from a different topological point in the phase diagram, green square in (a), can produce a more complicated behavior, whereby the Majorana is not destroyed after the minigap is reopened. {Instead, two new zero modes are added at a gap inversion, which takes place away from the high-symmetry $k=0$ point. Such kind of inversions are trivial, and introduce zero modes in pairs that hybridize to finite energy}; see Figs. 3 (e,f). Such split resonances are also eventually destroyed at higher mixing. 

Finally, a quite different scenario can take place. If $\eta$ is zero within the bulk of the nanowire, or due to some symmetry some of the $m_j\neq 0$ modes remain ungapped, the Majorana may become coupled to such gapless states by a local mixing $\eta$ confined to the tip of the nanowire, where the Majorana wavefunction is concentrated. Such a local mode mixing is a likely occurrence in experimental devices, since the tip of the nanowire is not covered by a superconducting shell, and is therefore more susceptible to mode-mixing perturbations from the substrate or tunnel probes. The result of such a local $\eta$ is shown in Fig. \ref{fig:4}. The background LDOS does not develop a minigap. Instead, the zero mode becomes broadened into a quasibound Majorana state, with a width that grows with $\eta$, and which represents its decay rate into the gapless nanowire bulk.

All these results assume a semi-infinite nanowire, without any longitudinal quantization of the different $m_j$ subbands. For finite nanowires the phenomenology becomes even more complicated, although in such a case one can no longer rigorously speak about topological nontriviality (at least in closed systems~\cite{Avila:CP19}). The general conclusion {from the analysis of our simple mode-mixing model}, however, is that while a small amount of mixing can be beneficial to stabilize the $m_j=0$ Majorana, it eventually leads to its destruction, either by broadening, splitting or a minigap closing and reopening. {We expect this qualitative behavior to be generic also in more elaborate models.}

\bibliography{biblio}

\end{document}